\documentstyle[fleqn,12pt,epsf]{article}
\def\pj{\hspace{-.26cm}}
\def\fpj{\hspace{-.7cm}}
\def\thalf{{\textstyle{\frac{1}{2}}}}
\def\frthtw{{\textstyle{\frac{3}{2}}}}
\def\frftw{{\textstyle{\frac{5}{2}}}}
\def\frstw{{\textstyle{\frac{7}{2}}}}

\def\ttwothird{{\textstyle{\frac{2}{3}}}}
\def\tquar{{\textstyle{\frac{1}{4}}}}
\def\toneig{{\textstyle{\frac{1}{8}}}}
\def\ttquar{{\textstyle{\frac{3}{4}}}}

\def\s{\sigma}
\def\so{\sigma_0}
\def\pv{\vmg{\pi}}

\def\epa{\epsilon_1'}

\newcommand{\vmg}[1]{\mbox{\boldmath$#1$}}
\begin{document}
\title{An Effective Lagrangian with Broken Scale and Chiral Symmetry IV:
\protect\\  Nucleons and Mesons at Finite Temperature}
\author{G.W. Carter$^\dagger$ and P.J. Ellis \\[2mm]
{\small School of Physics and Astronomy}\\[-2mm] {\small University of 
Minnesota, Minneapolis, MN 55455}\\[-1cm]}
\date{~}
\maketitle
\centerline{{\small\bf Abstract}}
\vskip-.15cm
{\small We study the finite temperature properties of an effective chiral 
Lagrangian which describes nuclear matter. Thermal fluctuations
in both the nucleon and the meson fields are considered. The 
logarithmic and square root terms in the effective potential are 
evaluated by expansion and resummation with the result written in 
terms of the exponential integral and the error function, respectively.
In the absence of explicit chiral symmetry breaking a phase transition 
restores the symmetry, but when the pion has a mass the transition is smooth.
The nucleon and meson masses as a functions of density and 
temperature are discussed.
\hfill\\
PACS: 11.10.Wx, 12.39.Fe, 24.10.Jv\hfill\\
Keywords: effective Lagrangian, finite temperature, chiral symmetry 
restoration\hfill\\
\hrule width 6cm\hfill\\
$^\dagger$ Present address: Niels Bohr Institute, 
University of Copenhagen, Denmark.
\thispagestyle{empty}
\vskip-20cm
\hfill NUC-MINN-97/7-T\\
\newpage
\section{Introduction}

In previous work \cite{glu3,glu4}, hereinafter referred to as I and II 
respectively, we have 
described an effective Lagrangian 
which incorporates broken scale symmetry in 
addition to spontaneously broken $SU(2)$ chiral symmetry, as suggested 
by quantum chromodynamics (QCD). This Lagrangian contained a potential 
with logarithmic 
terms involving the glueball field $\phi$ and the chiral $\sigma$ 
and $\pi$ fields. At temperature $T=0$ this led to a good description 
of nuclear matter and finite nuclei at the mean 
field level as well as low energy $\pi N$ 
scattering data. The extension of this type of model to $SU(3)$ has 
recently been discussed by Papazoglou {\it et al.} \cite{papa3}.
Here we examine the predictions of our $SU(2)$ Lagrangian for
the finite temperature, $T>0$, properties of hadronic matter which are 
needed in astrophysical applications and in the study of relativistic 
heavy ion collisions. Previous studies \cite{kal,papa} 
of models of this general type at $T>0$ have simply included temperature 
effects for the nucleons. Clearly thermal effects for the mesons
are also significant, particularly those due to the pion which will be 
dominant at low temperatures. 

The analysis at finite temperature is not straightforward due to
the logarithmic terms in the potential. We have recently suggested
in Ref. \cite{glu5}, hereinafter referred to as III, a method of treating 
these terms which involves expansion and 
resummation with the result cast in terms of the exponential integral.
This technique was applied to the meson sector of the effective Lagrangian.
Here we want to carry out a more complete calculation by including 
nucleons as well. As we shall show, this involves a square root term 
which, after analogous treatment, gives results which can be written in 
terms of the error function.
The Lagrangian and our thermal analysis are discussed 
in Section 2, with the detailed expressions for the necessary thermal 
averages being relegated to the Appendix.
We give our numerical results in Section 3 and Section 4 
contains our conclusions. 

\section{\bf Theory}
\subsection{\it Equations of Motion}

We will simplify the Lagrangian given in II by excluding 
the isotriplet vector and axial vector mesons, the $\rho$ and $a_1$.
They give no mean field contribution to symmetric nuclear matter and,
since they are relatively heavy, their thermal fluctuations will not 
play a significant role in chiral symmetry restoration. We can exclude an
additional term which was introduced to obtain the physical value 
of the axial coupling constant, $g_A$, since it involves the quantity
$\bar{N}\vmg{\tau}N$
which will not contribute for symmetric nuclear matter. We also discard a
term $\epsilon_3\bar{N}N$, which explicitly breaks chiral symmetry, since 
in II tiny values of $\epsilon_3$ in the range $0$ to 
$-15$ MeV were preferred. An unfavored symmetry breaking term 
labelled $\epsilon_2$ is also omitted. Then our effective Lagrangian 
can be written
\begin{eqnarray}
{\cal L}&\pj=&\pj\thalf\partial_{\mu}\sigma\partial^{\mu}
\sigma+\thalf\partial_{\mu}\vmg{\pi}\cdot\partial^{\mu}\vmg{\pi}
+\thalf\partial_{\mu}\phi\partial^{\mu}\phi
-\tquar\omega_{\mu\nu}\omega^{\mu\nu}
+\thalf G_{\omega\phi}\phi^2 \omega_\mu\omega^\mu 
\nonumber\\
&&+[(G_4)^2\omega_\mu\omega^\mu]^2
+\bar{N}\left[\gamma^\mu(i\partial_{\mu}-g_\omega\omega_\mu)
-g\sqrt{\s^2+\pv^2} \right] N -{\cal V}\,, \label{lb}\nonumber\\
{\cal V}&\pj=&\pj B\phi^4\hspace{-.7mm}
\left(\ln\frac{\phi}{\phi_0}-\frac{1}{4}\right)
\hspace{-.73mm}-\hspace{-.73mm}\thalf B\delta\phi^4
\ln\frac{\sigma^2+\vmg{\pi}^2}{\sigma_0^2}
+\hspace{-.74mm}\thalf B\delta \zeta^2\phi^2\!\!\left[\sigma^2
+\vmg{\pi}^2-\frac{\phi^2}{2\zeta^2}\right]\nonumber\\
&&-\tquar\epsilon_1'\left(\frac{\phi}{\phi_0}\right)^{\!2}
\left[\frac{4\sigma}{\sigma_0}-2\left(\frac{\sigma^2
+\vmg{\pi}^2}{\sigma_0^2}\right)-\left(\frac{\phi}{\phi_0}\right)^{\!2}
\,\right]-\ttquar\epsilon_1'\;. \label{lm}
\end{eqnarray}
Here $\zeta=\frac{\phi_0}{\sigma_0}$ and in the vacuum $\phi=\phi_0$, 
$\sigma=\sigma_0$ and $\vmg{\pi}=0$, regardless of whether or not the 
explicit symmetry breaking term $\epsilon_1'$ is present. The field 
strength tensor for the $\omega$ field is 
$\omega_{\mu\nu}=\partial_{\mu}\omega_{\nu}-\partial_{\nu}\omega_{\mu}$.
The $\omega$ mass is generated by coupling to the glueball field as 
was previously found to be necessary in order to describe nuclei. The
mass term can be written in terms of the vacuum mass 
$m_{\omega}=G_{\omega\phi}^{1/2}\phi_0$. In Eq. (\ref{lm}) we have also 
included a quartic term in the $\omega$ field since this generally
improved the phenomenology in I and II.

In Eq. (\ref{lb}) we have replaced the traditional coupling of the 
pion and the sigma meson to the nucleon, 
$-\bar Ng(\sigma+i\vmg{\pi}\cdot\vmg{\tau}\gamma_5)N$,
by $-\bar Ng\sqrt{\s^2+\pv^2}N$, which is a more natural form for including 
thermal fluctuations. As 
pointed out by Weinberg \cite{wein} such a transformation can be achieved 
by a redefinition of the nucleon field. The form of the $\omega$ coupling
is invariant, while the extra terms generated by the derivative involve
$\bar N\vmg{\tau}N$ and so will not contribute here. (The transformation 
has no effect if the pion field is set to zero as appropriate to $T=0$.)

The quantities $B$ and $\delta$ in Eq. (\ref{lm}) are parameters.
For the latter, guided by the QCD beta function, we take $\delta=4/33$ 
as in previous work. The logarithmic terms 
contribute to the trace anomaly: in addition to the standard contribution
from the glueball field \cite{mig,gomm} there is a contribution from 
the $\sigma$ field. There is also a contribution  from the explicit 
symmetry breaking, which is related to the pion mass.
Specifically the trace
of the ``improved" energy-momentum tensor is
$\theta_\mu^\mu=4\epsilon_{\rm vac}(\phi/\phi_0)^4$, where
the vacuum energy
$\epsilon_{\rm vac}=-\tquar B\phi^4_0(1-\delta)-\epsilon_1'$.

We take the vacuum glueball mass to be approximately 1.5 GeV
in view of QCD sum rule estimates \cite{shif} of 1.5 GeV and recent lattice 
estimates \cite{svw} of 1.7 GeV; small shifts in the precise value 
of the mass are inconsequential here.
Since the mass is large in comparison to the temperatures of interest,
we shall neglect thermal effects for the glueball.
We define the ratio of the mean field to the vacuum 
value to be $\chi=\phi/\phi_0$. Then 
the thermal averages of Lagrange's equations 
for the $\phi$, $\sigma$ and $\omega$ fields in uniform matter are:
\begin{eqnarray}
&&\fpj0=4B_0\chi^3\ln\chi-B_0\delta\chi\left\langle2\chi^2\ln\left(
\frac{\sigma^2+\vmg{\pi}^2}{\sigma_0^2}\right) - \left(
\frac{\sigma^2+\vmg{\pi}^2}{\sigma_0^2}\right)\right\rangle
-B_0\delta\chi^3\nonumber\\
&&\qquad-\epsilon_1'\chi\left\langle\frac{2\sigma}{\sigma_0}-\left(
\frac{\sigma^2+\vmg{\pi}^2}{\sigma_0^2}\right)\right\rangle
+\epsilon_1'\chi^3-m_{\omega}^2\chi
\langle\omega_{\mu}\omega^{\mu}\rangle\;,\nonumber\\[2mm]
&&\fpj0=-B_0\sigma_0^2\delta\chi^4\!\left\langle
\frac{\sigma}{\sigma^2+\vmg{\pi}^2}\right\rangle
\!+(B_0\delta\chi^2+\epsilon_1')\langle\sigma\rangle
-\epsilon_1'\chi^2\sigma_0+\left\langle\frac{g\sigma_0^2\sigma}
{\sqrt{\sigma^2+\vmg{\pi}^2}}\bar NN\right\rangle\!,\nonumber\\[2mm]
&&\fpj0=-m_\omega^2\chi^2\langle\omega_\mu\rangle
-4G_4^4\langle\omega_\nu\omega^\nu\omega_\mu\rangle
+g_\omega\langle\bar N \gamma_\mu N\rangle \;,\label{leq1}
\end{eqnarray}
where we have defined $B_0=B\phi_0^4$. Here the thermal averages are 
denoted by angle brackets.

Consider first thermal effects for the $\sigma$ and 
$\vmg{\pi}$ fields. We break $\sigma$ into a mean 
field part $\bar{\sigma}$ and a fluctuation $\Delta\sigma$ with mean 
value $\langle\Delta\sigma\rangle=0$. The mean value of the pion field 
$\langle\vmg{\pi}\rangle$ is, of course, zero. We write
\begin{eqnarray}
\frac{\sigma^2+\vmg{\pi}^2}{\sigma_0^2}&\pj=&\pj\frac{1}{\sigma_0^2}(
\bar{\sigma}^2+2\bar{\sigma}\Delta\sigma+\Delta\sigma^2
+\vmg{\pi}^2)\nonumber\\
&\pj\equiv&\pj \nu^2+2\nu\Delta\nu+\vmg{\psi}^2\;,\label{sigfl}
\end{eqnarray}
where, as in III, $\nu=\bar{\sigma}/\sigma_0$,
$\Delta\nu=\Delta\sigma/\sigma_0$ and 
$\vmg{\psi}^2=(\Delta\sigma^2+\vmg{\pi}^2)/\sigma_0^2$.
Since expanding the 
fluctuations out to lowest order will not 
properly treat $(\sigma^2+\vmg{\pi}^2)$ when it occurs in the 
denominator or in the logarithm in Eqs. (\ref{leq1}), we proceed as 
in III by expanding out the cross term $2\nu\Delta\nu$ of 
Eq. (\ref{sigfl}). The motivation is 
that at low temperatures the thermal average of 
$\vmg{\psi}^2$ is small, while at high temperatures $\nu$
is small, so the cross term is relatively small in both limits.
Note that odd powers of $\Delta\nu$ can be 
dropped since the thermal average gives zero. We also have to contend 
with products of meson and baryon factors. For these we adopt a 
factorization ansatz, {\it i.e.}, 
$\langle f(\s,\pv) \bar N N \rangle \simeq \langle f(\s,\pv)
\rangle\langle\bar N N\rangle$. For the omega field we write
$\omega_\mu = \omega_0 + \Delta\omega_\mu$ and terms linear in 
$\Delta\omega_\mu$ will vanish. The $G_4$ term, which involves four 
$\omega$ fields, will be treated at mean field level only, since it 
is a small correction and since this avoids the complication of 
differing longitudinal and transverse masses. With these approximations 
Eqs. (\ref{leq1}) become
\begin{eqnarray}
&&\fpj0=B_0\delta\chi\left\langle-2\chi^2\ln(\nu^2+\vmg{\psi}^2)
+\frac{4\chi^2\nu^2\Delta\nu^2}{(\nu^2+\vmg{\psi}^2)^2}
+\frac{8\chi^2\nu^4\Delta\nu^4}{(\nu^2+\vmg{\psi}^2)^4}+\vmg{\psi}^2
\right\rangle+B_0\delta\chi\nu^2\nonumber\\
&&\!\!\!\!+B_0\chi^3(4\ln\chi-\delta)-\epsilon_1'\chi(2\nu-\nu^2
-\langle\vmg{\psi}^2\rangle-\chi^2) -m_\omega^2\chi(\omega_0^2
+\langle\Delta\omega_\mu\Delta\omega^\mu\rangle)
,\nonumber\\[2mm]
&&\fpj0=B_0\delta\chi^4\nu\left\langle-\frac{1}{\nu^2+\vmg{\psi}^2}+
\frac{2\Delta\nu^2}{(\nu^2+\vmg{\psi}^2)^2}
-\frac{4\nu^2\Delta\nu^2}{(\nu^2+\vmg{\psi}^2)^3}
+\frac{8\nu^2\Delta\nu^4}{(\nu^2
+\vmg{\psi}^2)^4}\right\rangle\nonumber\\
&&\!\!\!+g\so\nu\rho_S\left\langle \frac{1}{\sqrt{\nu^2
+\vmg{\psi}^2}}-\frac{\Delta\nu^2}{(\nu^2+\vmg{\psi}^2)^{3/2}} 
+\frac{3\nu^2\Delta\nu^2}{2(\nu^2+\vmg{\psi}^2)^{5/2}} 
-\frac{5\nu^2\Delta\nu^4}{2(\nu^2+\vmg{\psi}^2)^{7/2}} 
\right\rangle\nonumber\\
&&\!\!\!+B_0\delta\chi^2\nu-\epsilon_1'
\chi^2(1-\nu)\;,\nonumber\\[2mm]
&&\fpj0=-m_\omega^2\chi^2\omega_0 - 4G_4^4\omega_0^3 + g_\omega\rho
\;. \label{leq2}
\end{eqnarray}
Notice that there is only a solution with exact chiral symmetry, 
$\nu=0$, when there is no explicit symmetry breaking, $\epsilon_1'=0$.
For most purposes it is sufficient to truncate these equations one 
power lower, however in the absence of explicit 
symmetry breaking this has the disadvantage that
there is a small region near the
chiral phase transition where solutions cannot be obtained. 
The expansion parameter is of order 
$4\nu^2\langle\Delta\nu^2\rangle/(\nu^2+\langle\vmg{\psi}^2\rangle)^2$
which {\it a posteriori} we find to be $<0.05$ -- this is 
satisfactorily small. 

The evaluation of the thermal average of the logarithm and the 
integral powers of $(\nu^2+\vmg{\psi}^2)$ was discussed in detail in III. 
A formal expansion in $\vmg{\psi}^2$
was made. The thermal average of $(\vmg{\psi}^2)^n$ was then written in
terms of $\langle\vmg{\psi}^2\rangle^n$ using counting factors that assume
$\langle\psi_i^2\rangle$ is independent of $i$, that is
$\langle\Delta\sigma^2\rangle=\langle\pi_a^2\rangle$ where $\pi_a$ is a 
component of the pion field. This is a high temperature approximation 
which will be accurate when chiral symmetry is exactly or 
approximately restored or when thermal contributions are dominant.
Nevertheless, our expressions also yield the correct 
low temperature limit. This suggests that our approximation is 
reasonable, although we do not have a quantitative assessment of the 
errors involved in the intermediate region. 
The final step is to resum the series and the result can be
written in terms of the exponential integral. A similar procedure is 
followed for the half-integral powers for which the result can be 
expressed in terms of the error function. This is discussed in 
Appendix A and 
expressions are listed there for all the thermal averages needed in 
Eqs. (\ref{leq2}) and the equations below. Note that the $T=0$ 
limit of Eqs. (\ref{leq2}) yields the relativistic mean field 
expressions given in II.

The evaluation of $\langle\vmg{\psi}^2\rangle$ requires the thermal 
average of the square of a scalar or pseudoscalar field. The
standard results are
\begin{equation}
\langle\pi_a^2\rangle=\frac{1}{2\pi^2}\int\limits_0^\infty\,dk
\frac{k^2}{e_{\pi}}\frac{1}{e^{\beta e_{\pi}}-1}
\quad;\quad \langle\Delta\sigma^2\rangle=
\frac{1}{2\pi^2}\int\limits_0^\infty\,dk
\frac{k^2}{e_{\sigma}}\frac{1}{e^{\beta e_{\sigma}}-1}\;. \label{tep}
\end{equation}
Here $\beta=1/T$ is the inverse temperature and $\pi_a$ represents a 
component of the pion field. The energies,
$e_{\pi}=\sqrt{k^2+m_{\pi}^{*2}}$ and
$e_{\sigma}=\sqrt{k^2+m_{\sigma}^{*2}}$, depend upon $m_{\pi}^{*2}$
and $m_{\sigma}^{*2}$, respectively the effective pion and sigma
masses. Here the asterisks denote the finite temperature and/or 
density masses which will be calculated in the 
next subsection. For the vector 
$\omega$ the corresponding result is
\begin{equation}
\langle\Delta\omega_{\mu}\Delta\omega^{\mu}\rangle=
-\frac{3}{2\pi^2}\int\limits_0^\infty\,dk
\frac{k^2}{e_{\omega}}\frac{1}{e^{\beta e_{\omega}}-1}\;, \label{teo}
\end{equation}
with $e_{\omega}=\sqrt{k^2+m_{\omega}^{*2}}$.
The nucleon density and the scalar density in Eqs. (\ref{leq2}) are
\begin{eqnarray}
\rho &\pj=&\pj \frac{2}{\pi^2}\int\limits_0^\infty{\rm d}k\,k^2 
\left(\frac{1}{e^{\beta(E^*-\mu^*)}+1}
-\frac{1}{e^{\beta(E^*+\mu^*)}+1}\right)\,,\nonumber\\
\rho_S &\pj=&\pj \frac{2M^*}{\pi^2}
\int\limits_0^\infty{\rm d}k\,\frac{k^2}{E^*} 
\left(\frac{1}{e^{\beta(E^*-\mu^*)}+1}
+\frac{1}{e^{\beta(E^*+\mu^*)}+1}\right)\,,
\label{dens}
\end{eqnarray}
where the effective chemical potential $\mu^*=\mu-g_{\omega}\omega_0$ 
and the energy $E^* = \sqrt{k^2+M^{*2}}$ with the nucleon effective 
mass $M^*$ defined below. In calculating 
thermodynamic integrals, such as these, we find it convenient to make 
use of the numerical approximation scheme of Ref. \cite{scott}.

\subsection{\it Masses}

Consistent with our factorization hypothesis, we define the effective 
nucleon mass to be
\begin{eqnarray}
M^*&\pj=&\pj g\left\langle\sqrt{\sigma^2+\vmg{\pi}^2}
\right\rangle\nonumber\\
&\pj=&\pj 
g\sigma_0\left\langle\sqrt{\nu^2+\vmg{\psi}^2}-\frac{\nu^2\Delta\nu^2}
{2(\nu^2+\vmg{\psi}^2)^{3/2}}-\frac{5\nu^4\Delta\nu^4}
{8(\nu^2+\vmg{\psi}^2)^{7/2}}\right\rangle\;,
\end{eqnarray}
where we have truncated the expansion at the same order as the equation
of motion. When $T=0$, Eq. (A.5) indicates 
that the usual result $M^*=g\sigma_0\nu$ is obtained, with the vacuum 
nucleon mass, $M=g\sigma_0$, determining $g$ for a given value of 
$\sigma_0$. Since we consider thermal fluctuations, the nucleon mass
will not become zero when chiral symmetry is restored and 
$\nu\rightarrow0$. In fact,
\begin{equation}
\lim_{\nu\rightarrow 0}\frac{M^*}{M} =
\sqrt{\frac{9\pi\langle\vmg{\psi}^2\rangle}{32}} \,.
\end{equation}
In the case where $\epsilon_1'=0$ the sigma and pion are massless at the 
chiral restoration temperature $T_c$, so 
$\langle\vmg{\psi}^2\rangle=T_c^2/(3\sigma_0^2)$ and 
$M^*/M= 0.543T_c/\sigma_0$.

For the meson fields we define the effective mass at finite temperature 
in terms of the thermal average 
of the second derivative of the Lagrangian. This means that we only 
consider contributions arising from a single interaction vertex. 
Since the mixing between the glueball and the $\sigma$ meson is 
small, we neglect it here for simplicity. Specifically
\begin{eqnarray}
\sigma_0^2m_{\sigma}^{*2}&\pj=&\pj
-\sigma_0^2\left\langle\frac{\partial^2{\cal L}}
{\partial\Delta\sigma^2}\right\rangle=(B_0\delta+\epsilon_1')\chi^2  
+\left\langle-\frac{B_0\delta\chi^4\sigma_0^2}{\sigma^2+\vmg{\pi}^2}
+\frac{2B_0\delta\chi^4\sigma_0^2\sigma^2}{(\sigma^2+\vmg{\pi}^2)^2}
\right\rangle\nonumber\\
&&\qquad\qquad\qquad\quad+ 
g\so^2\rho_S\left\langle\frac{1}{\sqrt{\s^2+\pv^2}} -
\frac{\s^2}{(\s^2+\pv^2)^{3/2}} \right\rangle\,,\nonumber\\[2mm]
\sigma_0^2m_{\pi}^{*2}&\pj=&\pj-\sigma_0^2\left\langle\frac{\partial^2
{\cal L}}{\partial\pi_a^2}\right\rangle=(B_0\delta+\epsilon_1')\chi^2
+\left\langle-\frac{B_0\delta\chi^4\sigma_0^2}{\sigma^2+\vmg{\pi}^2}
+\frac{2B_0\delta\chi^4\sigma_0^2\pi_a^2}{(\sigma^2+\vmg{\pi}^2)^2}
\right\rangle\nonumber\\
&&\qquad\qquad\qquad\quad+ g\so^2\rho_S\left\langle\frac{1}
{\sqrt{\s^2+\pv^2}}-
\frac{\pi_a^2}{(\s^2+\pv^2)^{3/2}} \right\rangle\,,\nonumber\\[2mm]
m_\omega^{*2}&\pj=&\pj 
\left\langle\frac{\partial^2{\cal L}}{\partial\Delta\omega_\mu
\partial\Delta\omega^\mu}\right\rangle = m_\omega^2\chi^2
\,,\nonumber\\[2mm]
\phi_0^2m_{\phi}^{*2}&\pj=&\pj-\phi_0^2\left\langle
\frac{\partial^2{\cal L}}
{\partial\phi^2}\right\rangle=4B_0\chi^2(3\ln\chi+1)+3
(\epsilon_1'-B_0\delta)\chi^2+(B_0\delta+\epsilon_1')\nu^2\nonumber\\
&&\qquad\qquad\qquad\quad-2\epsilon_1'\nu+\left\langle-6B_0\delta\chi^2
\ln\left(\frac{\sigma^2+\vmg{\pi}^2}{\sigma_0^2}\right)
+(B_0\delta+\epsilon_1')\vmg{\psi}^2\right\rangle\nonumber\\
&&\qquad\qquad\qquad\quad-m_\omega^2(\omega_0^2+
\langle\Delta\omega_\mu\Delta\omega^\mu\rangle)\;.\label{mass1}
\end{eqnarray}
Notice that at zero temperature a nucleon contribution to the pion 
mass of $g\sigma_0\rho_S/\nu$ is automatically obtained. In II 
this required the evaluation of the nucleon 
loop contribution to the pion propagator, whereas here it arises from 
the form of the coupling used for the chiral meson fields and the 
nucleon. As we have remarked, we do not consider thermal fluctuations 
in the glueball field and so its mass does not enter the equations. 
However it will be useful to display the mass in Sec. 3. The 
$\sigma$ and $\pi$ masses are needed in evaluating 
$\langle\vmg{\psi}^2\rangle$
and the $\omega$ mass depends on $\chi$. As a result, the equations of 
motion (\ref{leq2}) and the expressions for the masses (\ref{mass1}) 
must be evaluated self-consistently.

We need to expand the denominators in Eqs. (\ref{mass1}) as 
discussed in the previous subsection.
The expressions can be simplified by using 
the equations of motion (\ref{leq2}) 
for the case where $\nu\neq0$ and $\chi\neq0$, as well as the relation
\begin{equation}
\frac{1}{\sigma_0}\left\langle\frac{\sigma}{(\sigma^2
+\vmg{\pi}^2)^{\alpha}}
\right\rangle=\nu\left\langle\frac{1}{(\sigma^2+\vmg{\pi}^2)^{\alpha}}
\right\rangle-2\alpha\nu\left\langle\frac{\pi_a^2}
{(\sigma^2+\vmg{\pi}^2)^{\alpha+1}}
\right\rangle\;,
\end{equation}
which is valid in our approximation scheme for $\alpha$ an arbitary 
integer or half-integer. We obtain
\begin{eqnarray}
\sigma_0^2m_{\sigma}^{*2}&\pj=&\pj
2B_0\delta\chi^4\nu^2\!\left\langle\frac{1}{(\nu^2+\vmg{\psi}^2)^2}
-\frac{8\Delta\nu^2}{(\nu^2+\vmg{\psi}^2)^3}
+\frac{4(3\nu^2\Delta\nu^2+2\Delta\nu^4)}{(\nu^2+\vmg{\psi}^2)^4}\!
\right\rangle\!+\!\frac{\epsilon_1'\chi^2}{\nu}\nonumber\\
&&\!\!- g\so\rho_S\nu^2\left\langle\frac{1}{(\nu^2+\vmg{\psi}^2)^{3/2}}
-\frac{6\Delta\nu^2}{(\nu^2+\vmg{\psi}^2)^{5/2}}
+\frac{5(3\nu^2\Delta\nu^2+2\Delta\nu^4)}{2(\nu^2+\vmg{\psi}^2)^{7/2}}
\right\rangle , \nonumber\\[2mm]
\sigma_0^2m_{\pi}^{*2}&\pj=&\pj\frac{\epsilon_1'\chi^2}{\nu}
\;,\nonumber\\
\phi_0^2m_{\phi}^{*2}&\pj=&\pj4(B_0\chi^2+\epsilon_1'\nu)
-2(B_0\delta+\epsilon_1')(\nu^2+\langle\vmg{\psi}^2\rangle)
+ 2m_\omega^2(\omega_0^2+\langle\Delta\omega_\mu
\Delta\omega^\mu\rangle) \,.  \nonumber\\
&&\label{mass2}
\end{eqnarray}
It is straightforward to verify 
that in the zero temperature limit the results of II are obtained.
The explicit symmetry breaking parameter 
$\epsilon_1'=(\sigma_0m_\pi)^2$ is fixed by the 
vacuum pion mass and the chosen value of $\sigma_0$.
For the case when $\nu\rightarrow0$ the 
$\sigma$ and $\pi$ masses become equal:
\begin{equation}
\sigma_0^2m_{\sigma}^{*2}\rightarrow\sigma_0^2m_{\pi}^{*2}
\rightarrow B_0\delta\chi^2
\left(1-\frac{\chi^2}{\langle\vmg{\psi}^2\rangle}\right)
+g\so\rho_S\sqrt{\frac{9\pi}{32\langle\vmg{\psi}^2\rangle}} 
+\epsilon_1'\chi^2\;.\label{mnu0}
\end{equation}
When $\epsilon_1'=0$, the expression on the right is valid
for temperatures at which $\nu$ is exactly zero.

\subsection{\it Thermodynamics}

The grand potential per unit volume can easily be written down:
\begin{eqnarray}
\frac{\Omega}{V}&\pj=&\pj\langle{\cal V}\rangle
-\thalf m_\omega^{*2}\chi^2\omega_0^2 - G_4^4\omega_0^4
-\thalf m_{\sigma}^{*2}\langle\Delta\sigma^2\rangle
-\thalf m_{\pi}^{*2}\langle\vmg{\pi}^2\rangle\nonumber\\
&&+\frac{T}{2\pi^2}\int dk\,k^2
\left[\ln(1-e^{-\beta e_{\sigma}})+3\ln(1-e^{-\beta e_{\pi}})
+3\ln(1-e^{-\beta e_\omega})\right]\nonumber\\
&&-\frac{2T}{\pi^2}\int{\rm d}k\,k^2\left[
\ln\left(1+e^{-\beta(E^*-\mu^*)}\right)
+ \ln\left(1+e^{-\beta(E^*+\mu^*)}\right)\right] \;.\label{grand}
\end{eqnarray}
The subtraction of the fourth and fifth terms on the right
in Eq. (\ref{grand}) is 
necessary to avoid double counting \cite{lee}. 

Now if one takes the partial derivative of $\Omega/V$ with respect to
$\chi$, $\nu$, or $\omega_0$ the equations of motion (\ref{leq1}) 
are obtained. This is an important and non-trivial consistency check.
In order to show this one needs the equivalences
\begin{eqnarray}
\frac{\partial}{\partial\langle\vmg{\psi}^2\rangle} 
\left\langle \ln\left(\frac{\s^2+\pv^2}{\so^2}\right) \right\rangle 
&\pj=&\pj\frac{1}{2}\left\langle\frac{\so^2}{\s^2+\pv^2}\right\rangle
\;,\nonumber\\
\frac{\partial}{\partial\langle\vmg{\psi}^2\rangle} 
\left\langle \frac{\sqrt{\s^2+\pv^2}}{\so} \right\rangle &\pj=&\pj
\frac{3}{8}\left\langle\frac{\so}{\sqrt{\s^2+\pv^2}}\right\rangle \;.
\label{deriv}
\end{eqnarray}
We have not succeeded in proving these relations in general, but by 
expanding out $2\nu\Delta\nu$ as before and using the explicit 
equations in the Appendix we have verified Eqs. (\ref{deriv}) to 
orders $(\nu^2+\vmg{\psi}^2)^{-4}$ and 
$(\nu^2+\vmg{\psi}^2)^{-9/2}$, respectively, in our approximation 
scheme. This is all that is needed here. Furthermore
Eqs. (\ref{deriv}) are
exact in the limits of zero and infinite temperature.

In practice it is necessary to truncate the expansions.
If the equations for the fields, the masses and the grand potential
are truncated at the same order,
then the pion mass is exactly zero in the absence of explicit
symmetry breaking and there is approximate consistency between 
the equations of motion and the grand potential. Alternatively
the equations can be truncated at different orders such that there is 
exact consistency between the equations of motion and the grand 
potential, but then the pion mass is non-zero at low temperature
when $\epsilon_1'=0$, thus violating Goldstone's 
theorem. We choose the former alternative and note that the inaccuracy 
is not large, although it can result in a small negative value for 
the pressure at low temperature.

We therefore truncate the expansion for the thermal average of the 
potential at order $(\nu^2+\vmg{\psi}^2)^{-4}$ and write
\begin{eqnarray}
\langle{\cal V}\rangle&\pj=&\pj\chi^4[B_0\ln\chi-\tquar 
B_0(1+\delta)+\tquar\epsilon_1')] +\thalf(B_0\delta+\epsilon_1')
\chi^2(\nu^2+\langle\vmg{\psi}^2\rangle)\nonumber\\
&&-\epsilon_1'\chi^2\nu -\thalf B_0\delta\chi^4\left\langle
\ln(\nu^2+\vmg{\psi}^2)-\frac{2\nu^2\Delta\nu^2}
{(\nu^2+\vmg{\psi}^2)^2}-\frac{4\nu^4\Delta\nu^4}
{(\nu^2+\vmg{\psi}^2)^4}\right\rangle\nonumber\\
&&+\tquar[B_0(1-\delta)+\epsilon_1']\;.
\end{eqnarray}
We have added a constant term here so that $\langle{\cal V}\rangle$
is zero in the vacuum. The pressure $P$ is of course $-\Omega/V$.

It is straightforward to derive the energy density, which takes 
the form
\begin{eqnarray}
\frac{E}{V}&\pj=&\pj\langle{\cal V}\rangle
+\thalf m_\omega^{*2}\chi^2\omega_0^2 + 3 G_4^4 \omega_0^4
-\thalf m_{\sigma}^{*2}\langle\Delta\sigma^2\rangle
-\thalf m_{\pi}^{*2}\langle\vmg{\pi}^2\rangle\nonumber\\
&&+\frac{1}{2\pi^2}\int dk\,k^2\left[\frac{e_{\sigma}}
{e^{\beta e_{\sigma}}-1}+\frac{3e_{\pi}}{e^{\beta e_{\pi}}-1}
+\frac{3e_{\omega}}{e^{\beta e_{\omega}}-1}\right]\nonumber\\
&&+\frac{2}{\pi^2}\int {\rm d}k \, k^2E^*
\left(\frac{1}{e^{\beta(E^*-\mu^*)}+1}+\frac{1}
{e^{\beta(E^*+\mu^*)}+1}\right)\;.
\end{eqnarray}

\section{Results}

The parameters used to obtain the numerical results
are listed in Table 1. They were determined in II by fitting to 
the properties of nuclear matter and finite nuclei. Quite a reasonable 
phenomenology was obtained there, with the non-zero value of $G_4$ being
slightly favored, so it is sensible to explore the behavior at $T>0$.
We shall also consider the case where the vacuum pion mass vanishes 
($\epsilon_1'=0$) for which the parameters differ only slightly 
from those specified in Table 1.

\begin{table}[t]
\begin{center}
\caption{Values of the parameters.}
\begin{tabular}{|l|c|c|} \hline
Quantity& $G_4=0$ & $G_4/g_{\omega}=0.19$\\ \hline
$|\epsilon_{\rm vac}|^{1/4}$ (MeV)& 236 & 228\\
$g_{\omega}$& 10.5 & 12.2 \\
$\zeta=\phi_0/\sigma_0$& 1.28 & 1.41\\
$\sigma_0$ (MeV) & 110 & 102\\
${\epsilon_1'}^{1/4}$ (MeV)& 123 & 119\\ \hline
\end{tabular}
\end{center}
\end{table}

If $\epsilon_1'=0$, so that the pion is massless in the vacuum, 
chiral symmetry can be exactly restored at sufficiently high 
temperature. We show in Fig. 1 the chiral restoration temperature 
$T_c$ versus the ratio of baryon 
density $\rho$ to the density at saturation $\rho_0$.
The primary distinction between the zero and non-zero $G_4$ cases 
arises from the different values of $\so$ in Table 1, which
fix the overall temperature scale. Thus the same qualitative behavior 
is seen in the two cases with the corresponding temperatures slightly 
larger when $G_4=0$; at zero density this scaling
factor is almost exactly $110/102 \approx 1.1$, but it increases to 
about $1.7$ at very high density. Therefore it is sufficient in 
the following to focus on the $G_4\neq0$ results.

In III, baryons and $\omega$ mesons were absent and $T_c$ was 187 
MeV. Since at restoration $m_\sigma^*=m_\pi^*=0$, 
Eq. (\ref{mnu0}) shows that $T_c$ is modified by the scalar 
density which is non-zero even at zero baryon density. We obtain 
a value of 162 MeV, which is a reasonable 
order of magnitude in view of the many estimates in the literature. We 
mention that the standard Gell-Mann-L\'evy model \cite{sasha} yields 
$T_c=\sqrt{2}f_{\pi}=132$ MeV. Also Gerber and Leutwyler \cite{gerb} 
estimate $T_c=190$ MeV in two-flavor QCD using an effective chiral 
Lagrangian to three loop order, while a recent lattice 
calculation \cite{bern97}
gave $140-150$ MeV. The restoration temperature decreases with 
increasing density so that at $\rho_0$ it is 125 MeV and asymptotically 
it is 45 MeV. A finite $T_c$ is required in the high density limit
since at $T=0$, as shown in Fig. 2, 
$\nu$ becomes $\simeq0.3$, which is small but non-zero.
Thus chiral symmetry is not completely restored at zero temperature
in our model; this finds some support from lattice calculations 
\cite{blum}. Papazoglou {\it et al.} \cite{papa} have found
that there are regions in parameter space for models of this type
where chiral symmetry is restored at high density, but with these
potentials the compression modulus is unreasonably high.

The remainder of Fig. 2 shows that $\nu$ becomes zero at progressively 
lower densities as the temperature is raised. The dotted lines here 
indicate metastable regions and, while the transition is clearly first 
order, we do not believe that our approximation scheme is sufficiently 
accurate to reliably determine the order. For instance, if our expansions 
are truncated at a lower order we are unable to find solutions for a 
small region in the vicinity of the transition. Suppressing results in 
the metastable regions, we plot the sigma mass in Fig. 3. 
For $T=0$ the mass increases with density due to the scalar 
density contribution in Eq. (\ref{mass2}). With increasing temperature 
the familiar behavior is seen, namely, the mass drops to zero to 
become degenerate with the pion mass at successively lower densities.
Beyond this point the two masses increase together
as either the temperature or density is raised.

We now turn to the more realistic case where $\epsilon_1'\neq0$ and the 
pion has its physical mass in the vacuum. Figure 4 plots $\nu$ as a 
function of density. At low temperatures $\nu$ increases slightly 
for finite densities, before decreasing as the temperature is 
further raised and becoming 
rather independent of density. Substantial 
reduction in $\nu$ requires higher temperatures than needed for Fig. 2,
which seems reasonable in view of the finite pion mass here. 
With baryons present $\nu$ smoothly tends towards zero as 
$T\rightarrow\infty$, whereas in their absence it stabilized at
$\simeq$0.15, as demonstrated in III. In either case there is no 
phase transition when $\epsilon_1'\neq0$. In this connection it 
is worth mentioning that the lattice QCD calculations of Brown 
{\it et al.} \cite{brown} show that a second order phase transition 
for massless quarks is washed out when they are given a finite mass.

The pion and sigma masses are shown in Figs. 5 and 6. As the 
temperature is raised the pion mass increases, while the sigma mass 
decreases so that they approach each other. They are degenerate 
when a temperature of $\sim250$ MeV 
is reached and depend very little on density. Thereafter they continue 
to rise, passing the 1 GeV mark at $T\simeq 300$ MeV (not illustrated). 
The sigma mass remains above 400 MeV for all densities and temperatures.
This is qualitatively different from the high temperature behavior
for the $\epsilon_1'=0$ case shown in Fig. 3. Furthermore, comparison 
with III indicates that the presence of baryons pushes up the mass 
noticeably.

The nucleon effective mass, $M^*$, in shown in Fig. 7. At low 
temperatures its behavior is dominated by $\nu$ and the mass 
increases slightly with temperature for non-zero densities. 
However, the thermal fluctuations, $\langle\vmg{\psi}^2\rangle$, 
begin to dominate for temperatures above 150 MeV, leading to a drop 
in $M^*$ at low densities. This thermal domination implies that the 
behavior of the mean field $\nu$ is of minor consequence, which is why 
a plot of $M^*$ for the $\epsilon_1'=0$ case looks quite similar to 
Fig. 7. By $T=250$ MeV, $M^*$ is rather independent of density.
The mass continues to fall at higher temperatures 
and remains roughly independent of density.
At such high $T$ this is not caused by the very small $\nu$, but
rather by the diminished value of $\langle\vmg{\psi}^2\rangle$ which 
follows from the sharp increase in the meson masses seen in Figs. 5 and 6.
Qualitatively the overall behavior of $M^*$ is strikingly similar to that 
reported by Furnstahl and Serot \cite{fs} for the Walecka model. 
For a given $T$, their mass drops to a lower level than ours,
probably because they did not consider meson fluctuations.
On the other hand at least one model -- the quark meson coupling 
model --  shows a different behavior \cite{panda} in that the 
nucleon mass simply increases monotonically with temperature.

The ratio of the glueball mean field to the vacuum value, 
$\chi=\phi/\phi_0$, is shown in Fig. 8. This ratio differs from unity 
by more than 10\% only for the highest temperature, 250 MeV, and low 
to moderate densities. To complete the discussion, we must examine 
the glueball effective mass of Eq. (\ref{mass2}), $m_\phi^*$, which 
is displayed in Fig. 9.
Again we remark that this mass was not used in the self-consistent 
solution of the equations since glueball fluctuations were excluded 
on the grounds 
that the mass is large. This is proven justified for low temperatures,  
but it is becoming questionable at $T=250$ MeV for low density.
Also the difference between $m_{\phi}^*$ and $m_{\sigma}^*$ starts to 
become small and it may not be adequate to ignore the mixing induced by
the off-diagonal terms in the mass matrix described in II (the 
vacuum mass in Fig. 9 will be pushed up a little by this mixing).
Furthermore, beyond this temperature the fluctuations of the 
$\omega$ meson field,
$\langle\Delta\omega_\mu\Delta\omega^\mu\rangle$,
start to become significant.
This reduces the glueball effective mass further and continuing to 
solve for $m_\omega^*$ and $\chi$ self-consistently quickly sends
$m_\phi^*$ to zero at $T\sim 300$ MeV. This effect was not observed 
in III since the $\omega$ meson was not included. To obtain 
reliable solutions in this region one would first of all need to 
consider fluctuations in the glueball field and compute the mixing 
with the $\sigma$ field. Secondly, since the mass of the $\rho$ 
meson is similar to that of the $\omega$, the fluctuations in this 
field will also play a role even though its mean field is zero. 
In addition one should include
the chiral partner of the $\rho$, namely the $a_1$, since with the 
approximate restoration of chiral symmetry the masses will be similar. 
If the effect of fluctuations in the $\rho$ 
and $a_1$ fields is estimated by appropriately increasing the degeneracy 
factor for the $\omega$ fluctuations, it is found that $m_\phi^*$ 
is driven to zero at a temperature of $\sim$200 MeV (at zero density;
the effect is softened with increasing density, but still present). 
Although we expect that glueball fluctuations will partially counter this 
effect, it is intriguing since it 
could be an intimation of deconfinement. Lattice 
calculations suggest that this occurs at a similar temperature 
to chiral restoration 
\cite{satz}. However a consistent thermal analysis for the $\phi$, 
$\rho$ and $a_1$ mesons would be quite involved and is beyond 
the scope of the present work.

Turning to the thermodynamics of our model, we first remark that 
in common with other models we observe a liquid-gas phase 
transition; the critical temperature is found to be 16 MeV. 
The pressure and energy density are plotted in Figs. 10 and 11, 
respectively, as functions of density.  The calculations here are for
$\epa\neq0$, but taking $\epa=0$ yields similar results, except that
all numbers are reduced by a few percent in the absence of a finite 
pion mass. At the highest temperatures and densities shown, it can 
be noted that $E \sim 3PV$, which is the massless gas result. However 
this does not hold for lower values of either $T$ or $\rho$, for 
which interactions dominate the thermal fluctuations. For orientation 
we have compared the pressure to that obtained with a crude lowest 
order treatment of gluons and massless quarks, taking the bag 
constant to be $|\epsilon_{\rm vac}|$. We find that at no point, 
either in temperature or density, is the quark-gluon plasma the 
preferred phase 
prior to chiral restoration (for $\epsilon_1'=0$).  At zero density, 
for example, chiral restoration is achieved  at $T_c=162$ MeV while
the quark-gluon plasma has a larger pressure for $T\simeq170$ MeV.
At equilibrium density, where 
we found $T_c=125$ MeV, the quark-gluon phase is preferred 
at $T=164$ MeV.

\section{Conclusions}

We have discussed the finite temperature behavior of 
an effective Lagrangian with which we have successfully described 
nuclear matter and finite nuclei at $T=0$. With nucleons present, we 
have thermally averaged a square root term involving the meson 
fields in addition to 
a logarithmic term. The latter was handled in III by expansion and 
resummation of an infinite series with the final result cast in terms 
of the exponential integral. For the former the result of a similar 
approach was written in terms of the error function.

Our results showed that at sufficiently high temperature the mean 
value of the $\sigma$ field became small, signalling chiral restoration. 
In the absence of explicit chiral symmetry breaking, a phase transition
restored the symmetry at temperatures ranging from 162 MeV at zero 
density to 45 MeV at very large density. It was estimated that that 
the quark-gluon phase would be preferred at somewhat higher 
temperatures than these. In the physical case, where explicit chiral 
symmetry breaking was present and the pion had a vacuum mass, a 
smooth restoration of chiral symmetry was found.
The masses of the pion and $\sigma$ meson were virtually degenerate
and the order parameter $\nu$ became small for temperatures of 
$\sim250$ MeV.
At such a temperature the nucleon mass was also reduced, but the effect
of meson fluctuations yielded some stabilization so that 
$M^*/M\simeq0.43$.
At a similar temperature the glueball mass started to be 
significantly reduced due to the coupling to $\omega$ vector meson
fluctuations. This could be a hint of interesting physics, but we were 
unable to track it further with the present 
formalism since glueball fluctuations and the $\rho$ and $a_1$ mesons
are expected to be of significance.

We thank S. Rudaz for useful discussions. We acknowledge partial 
support from the Department of Energy under grant No. 
DE-FG02-87ER40328. G.W.C. thanks the University of Minnesota for a 
Doctoral Dissertation Fellowship. A grant for computing time from 
the Minnesota Supercomputer Institute is gratefully acknowledged.

\section*{\bf Appendix A. Thermal Averages}
\setcounter{equation}{0}
\renewcommand{\theequation}{A.\arabic{equation}}
\noindent{\it A.1. Half-Integral Powers}

We first write the expansion
\begin{equation}
\left\langle \sqrt{\nu^2+\vmg{\psi}^2}\,\right\rangle=
\nu\left\langle1+\frac{\vmg{\psi}^2}{2\nu^2}
-\sum_{n=2}^\infty(-1)^n\frac{(2n-3)!!}{2^nn!}
\frac{\vmg{\psi}^{2n}}{\nu^{2n}}\right\rangle\;.
\end{equation}
For the purposes of evaluating the counting, we assume that the 
thermal average $\langle\psi_i^2\rangle$ is independent of the 
label $i$, which amounts to 
assuming the masses of the particles involved are the same. 
Then the result of taking the thermal average 
of each possible pair of fields at a general vertex can be written
$\langle(\vmg{\psi}^2)^n\rangle=c_n\langle\vmg{\psi}^2\rangle^n$
and, as shown in III, $c_n=(n+1)!/2^n$. Using this result and defining
\begin{equation}
\frac{1}{z^2} = \frac{\langle\vmg{\psi}^2\rangle}{2\nu^2}\;, 
\end{equation}
we have
\begin{equation}
\left\langle\sqrt{\nu^2+\vmg{\psi}^2}\,\right\rangle =
\nu\left[ 1 + \frac{1}{z^2} + \sum_{m=1}^\infty (-1)^m
\frac{(m+2)(2m-1)!!}{2^{m+1}} z^{-2(m+1)} \right]\;.
\label{sqrtsum}
\end{equation}
This is a divergent series which we regard as a formal expansion 
and it must be resummed before it can be evaluated.
Matching this expression to the asymptotic expansions of 
${\rm i}^n{\rm erfc}(z)$, the repeated integrals of the complementary 
error function \cite{as}, we have
\begin{eqnarray}
\left\langle\sqrt{\nu^2+\vmg{\psi}^2}\,\right\rangle &\pj=&\pj
\nu + \frac{\nu}{4z}\sqrt{\pi}e^{z^2}
\left[2z{\rm i}^1{\rm erfc}(z)+3{\rm i}^0{\rm erfc}(z)\right]
\nonumber\\
&\pj=&\pj\frac{3\nu}{4z}\left[ 2z+(1-\ttwothird z^2)
\sqrt{\pi}e^{z^2}{\rm erfc}(z)\right]\;.
\end{eqnarray}
Here 
${\rm erfc}(z)=1-2\pi^{-\frac{1}{2}}\int_0^ze^{-t^2}dt$.
In the limit of low temperature ($z\rightarrow\infty$)
\begin{equation}
\left\langle\sqrt{\nu^2+\vmg{\psi}^2}\,\right\rangle\rightarrow
\nu+\frac{\langle\vmg{\psi}^2\rangle}{2\nu}\;,
\end{equation}
as follows from expanding out the square root directly. 
In the high temperature limit ($z\rightarrow0$)
\begin{equation}
\left\langle\sqrt{\nu^2+\vmg{\psi}^2}\,\right\rangle\rightarrow
\frac{3}{4}\sqrt{\frac{\pi}{2}}\sqrt{\langle\vmg{\psi}^2\rangle}\;,
\end{equation}
and the numerical factor is 0.9400 which is close to unity,
as one might expect.

The other expressions that are needed can be obtained by 
differentiating with respect to $\nu^2$ and by using the relations
\begin{eqnarray}
\left\langle\Delta\nu^2f(\nu,\vmg{\psi}^2)\right\rangle&\pj=&\pj
\tquar\left\langle\vmg{\psi}^2f(\nu,\vmg{\psi}^2)\right\rangle
\;,\nonumber\\
\left\langle\Delta\nu^4f(\nu,\vmg{\psi}^2)\right\rangle&\pj=&\pj
3\sigma_0^{-2}\left\langle\Delta\nu^2\pi_a^2f(\nu,\vmg{\psi}^2)
\right\rangle=
\toneig\left\langle\vmg{\psi}^4f(\nu,\vmg{\psi}^2)\right\rangle\;,
\end{eqnarray}
which are easily shown for an arbitrary function $f$, when 
$\langle\psi_i^2\rangle$ is independent of $i$. One obtains 
the following:
\begin{eqnarray}
&&\fpj\left\langle (\nu^2+\vmg{\psi}^2)^{-\thalf} \right\rangle =
\frac{z}{2\nu}\left[ 2z+(1-2z^2)\sqrt{\pi}e^{z^2}{\rm
erfc}(z)\right]\;,\nonumber\\
&&\fpj\left\langle (\nu^2+\vmg{\psi}^2)^{-\frthtw} \right\rangle =
-\frac{z^3}{\nu^3}\left[ 2z-(1+2z^2)\sqrt{\pi}e^{z^2}{\rm
erfc}(z)\right]\;,\nonumber\\
&&\fpj\left\langle \Delta\nu^2(\nu^2+\vmg{\psi}^2)^{-\frthtw} 
\right\rangle=\frac{z}{8\nu}\left[ 2z+4z^3+(1-4z^2-4z^4)
\sqrt{\pi}e^{z^2}{\rm erfc}(z)\right]\;,\nonumber\\
&&\fpj\left\langle \Delta\nu^2(\nu^2+\vmg{\psi}^2)^{-\frftw} 
\right\rangle=-\frac{z^3}{12\nu^3}\left[ 10z+4z^3-(3+12z^2+4z^4)
\sqrt{\pi}e^{z^2}{\rm erfc}(z)\right]\:,\nonumber\\
&&\fpj\left\langle \Delta\nu^2(\nu^2+\vmg{\psi}^2)^{-\frstw} 
\right\rangle=\frac{z^4}{30\nu^5}\Bigl[ 8+18z^2+4z^4\nonumber\\
&&\hspace{4.5cm} -(15z+20z^3+4z^5)\sqrt{\pi}e^{z^2}{\rm
erfc}(z)\Bigr]\;,\nonumber\\
&&\fpj\left\langle \Delta\nu^4(\nu^2+\vmg{\psi}^2)^{-\frstw} 
\right\rangle=-\frac{z^3}{120\nu^3}\Bigl[66z+56z^3+8z^5 \nonumber\\
&&\hspace{4.5cm}
- (15+90z^2+60z^4+8z^6)\sqrt{\pi}e^{z^2}{\rm erfc}(z)\Bigr].
\end{eqnarray}

\noindent{\it A.2. Integral Powers}

The thermal averages needed can be derived in the manner indicated 
above using the thermal average of the logarithm. This is evaluated 
in analogous fashion to the square root and the details were discussed 
in III. We simply list the results here, in terms of the 
variable $y=z^2$:
\begin{eqnarray}
&&\fpj\left\langle {\rm ln}(\nu^2+\vmg{\psi}^2) \right\rangle=
{\rm ln}\nu^2 + (1-y)e^yE_1(y) + 1 \;,\nonumber\\
&&\fpj\left\langle (\nu^2+\vmg{\psi}^2)^{-1} \right\rangle=
\frac{y}{\nu^2} \left[1 - y e^y E_1(y)\right] \;,\nonumber\\
&&\fpj\left\langle (\nu^2+\vmg{\psi}^2)^{-2} \right\rangle=
-\frac{y^2}{\nu^4} \left[1-(1+y)e^y E_1(y)\right] \;,\nonumber\\
&&\fpj\left\langle (\nu^2+\vmg{\psi}^2)^{-3} \right\rangle=
\frac{y^2}{2\nu^6} \left[1+y-(2y+y^2)e^y E_1(y)\right] 
\;,\nonumber\\
&&\fpj\left\langle \Delta\nu^2 (\nu^2+\vmg{\psi}^2)^{-2} 
\right\rangle=\frac{y}{4\nu^2} \left[1+y-(2y+y^2)e^y E_1(y)\right] 
\;,\nonumber\\
&&\fpj\left\langle \Delta\nu^2 (\nu^2+\vmg{\psi}^2)^{-3} 
\right\rangle=-\frac{y^2}{8\nu^4} \left[3+y-(2+4y+y^2)e^y 
E_1(y)\right] \;,\nonumber\\
&&\fpj\left\langle \Delta\nu^2 (\nu^2+\vmg{\psi}^2)^{-4} 
\right\rangle=\frac{y^2}{24\nu^6} \left[2+5y+y^2-(6y+6y^2+y^3) 
e^y E_1(y)\right] \;,\nonumber\\
&&\fpj\left\langle \Delta\nu^4 (\nu^2+\vmg{\psi}^2)^{-4} 
\right\rangle=
-\frac{y^2}{48\nu^4} \Bigl[11+8y+y^2\nonumber\\
&&\hspace{4.5cm}-(6+18y+9y^2+y^3)e^y E_1(y)\Bigr] \,. 
\end{eqnarray}
The exponential integral is defined \cite{as} by
$E_1(y)=\int_1^\infty dt\,t^{-1}e^{-yt}$.

\begin{figure}[t]
\setlength\epsfxsize{6.0in}
\centerline{\epsfbox{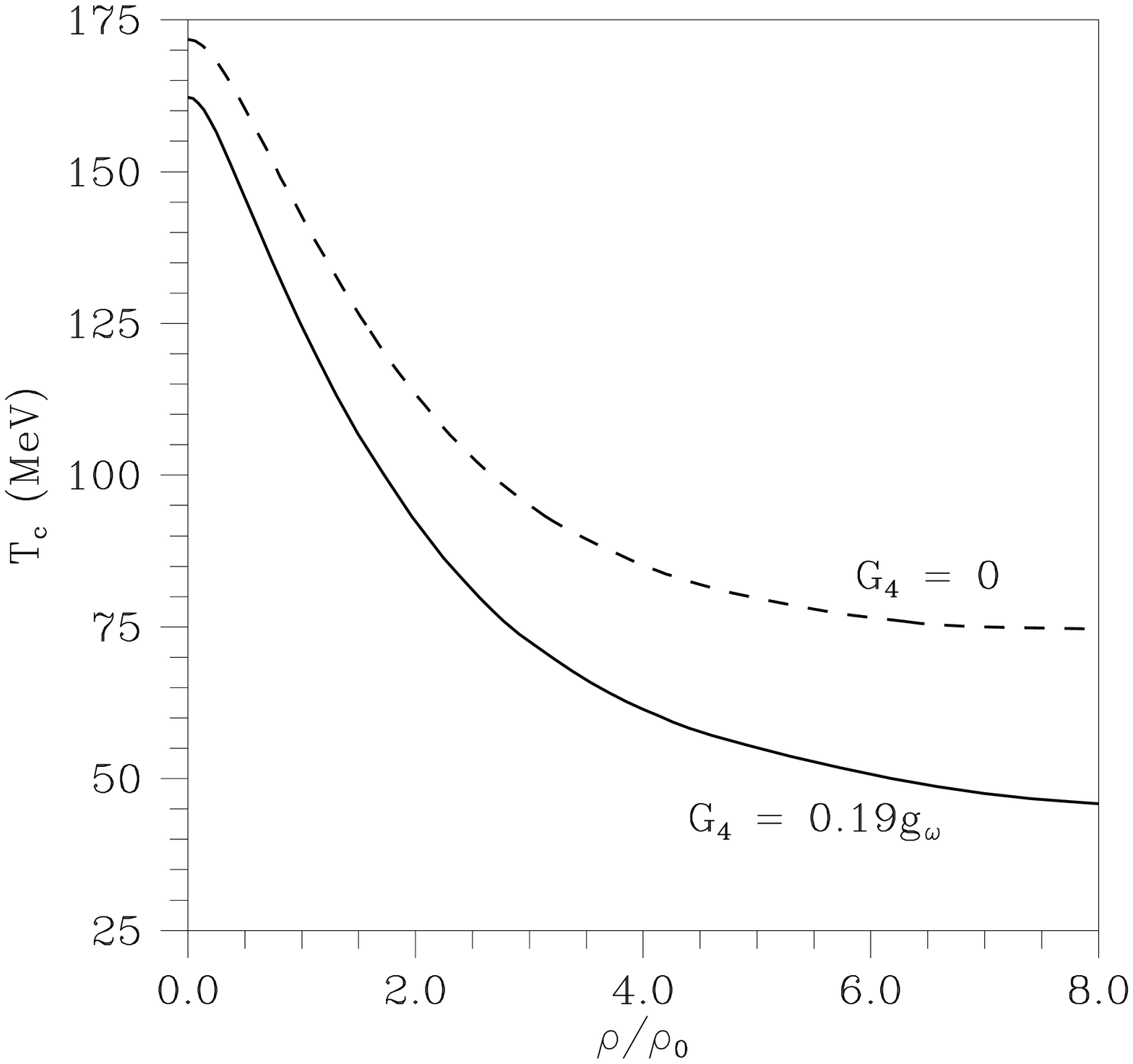}}
\noindent {\bf Figure 1.} The chiral restoration temperature 
$T_c$ as a function of the ratio of the density to equilibrium 
nuclear matter density, $\rho/\rho_0$, for two values of $G_4$ 
with $\epsilon_1'=0$.
\end{figure}

\begin{figure}[t]
\setlength\epsfxsize{6.0in}
\centerline{\epsfbox{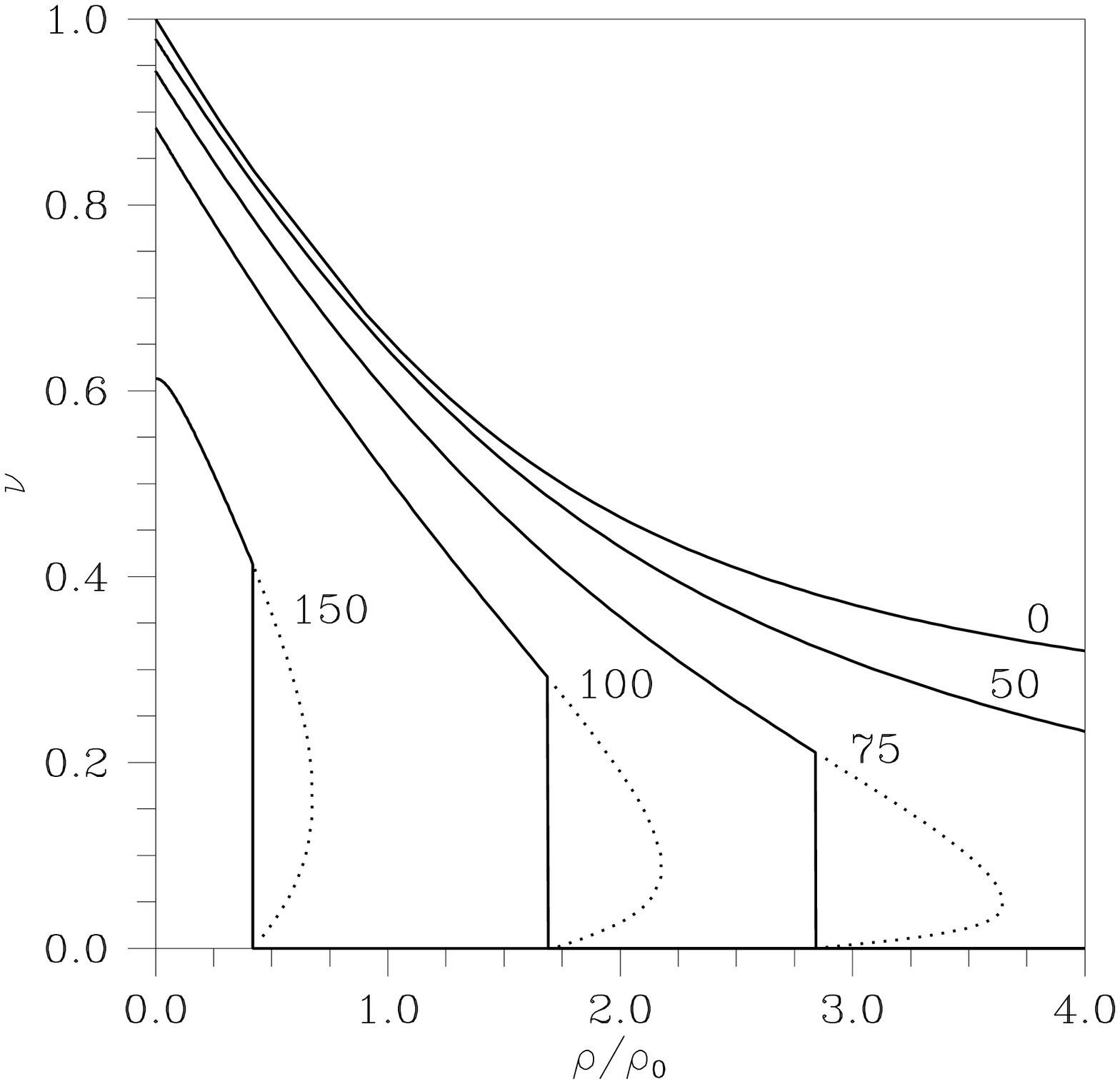}}
\noindent {\bf Figure 2.} The mean sigma field, 
$\nu=\bar{\sigma}/\sigma_0$, 
as a function of density for various temperatures (in MeV)
with $\epsilon_1'=0$. The dotted curves
indicate thermodynamically unstable regions.
\end{figure}

\begin{figure}[t]
\setlength\epsfxsize{6.0in}
\centerline{\epsfbox{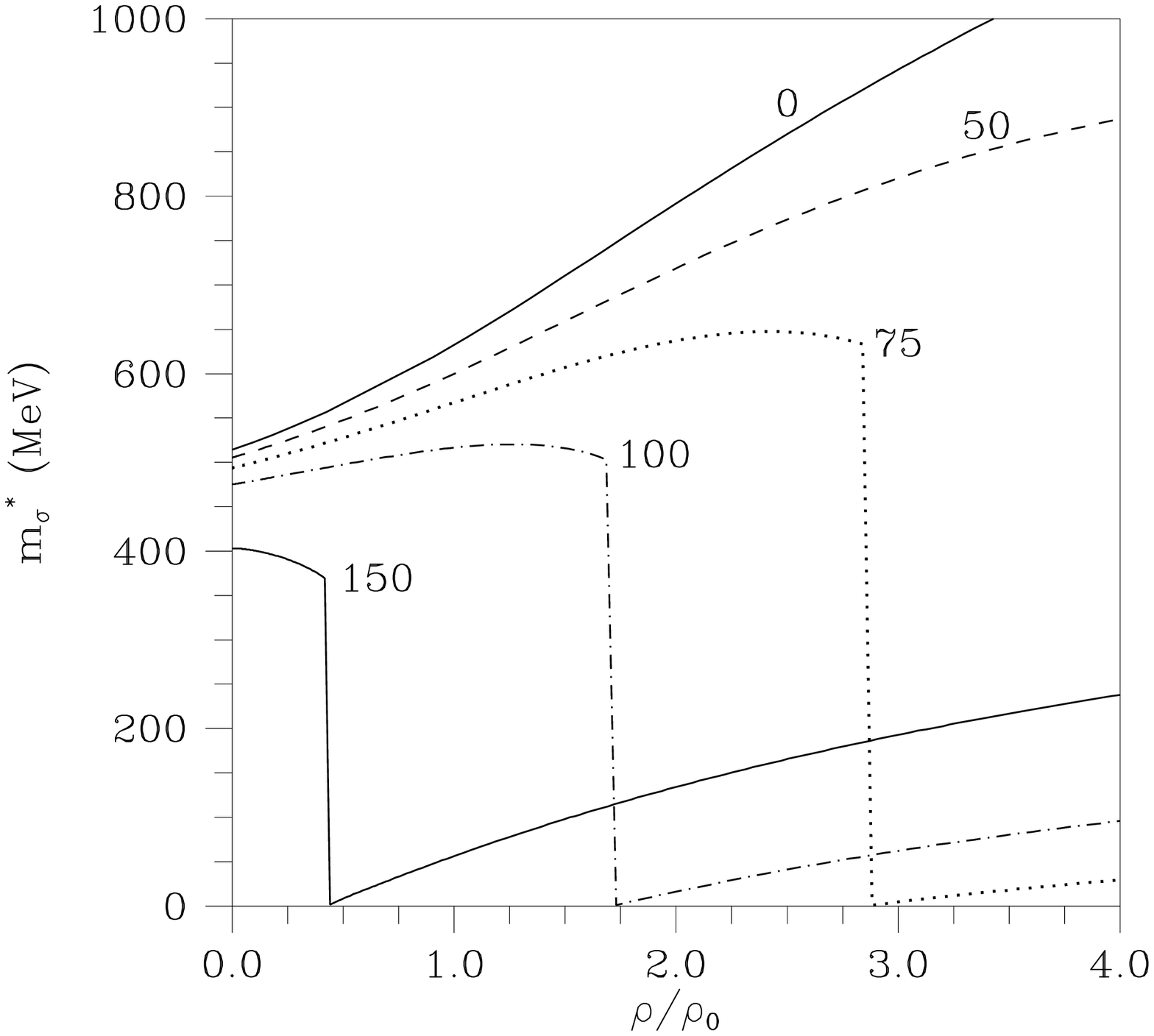}}
\noindent {\bf Figure 3.} The sigma effective mass, 
$m_{\sigma}^*$, as a function of density for various temperatures 
(in MeV) with $\epsilon_1'=0$. 
\end{figure}

\begin{figure}[t]
\setlength\epsfxsize{6.0in}
\centerline{\epsfbox{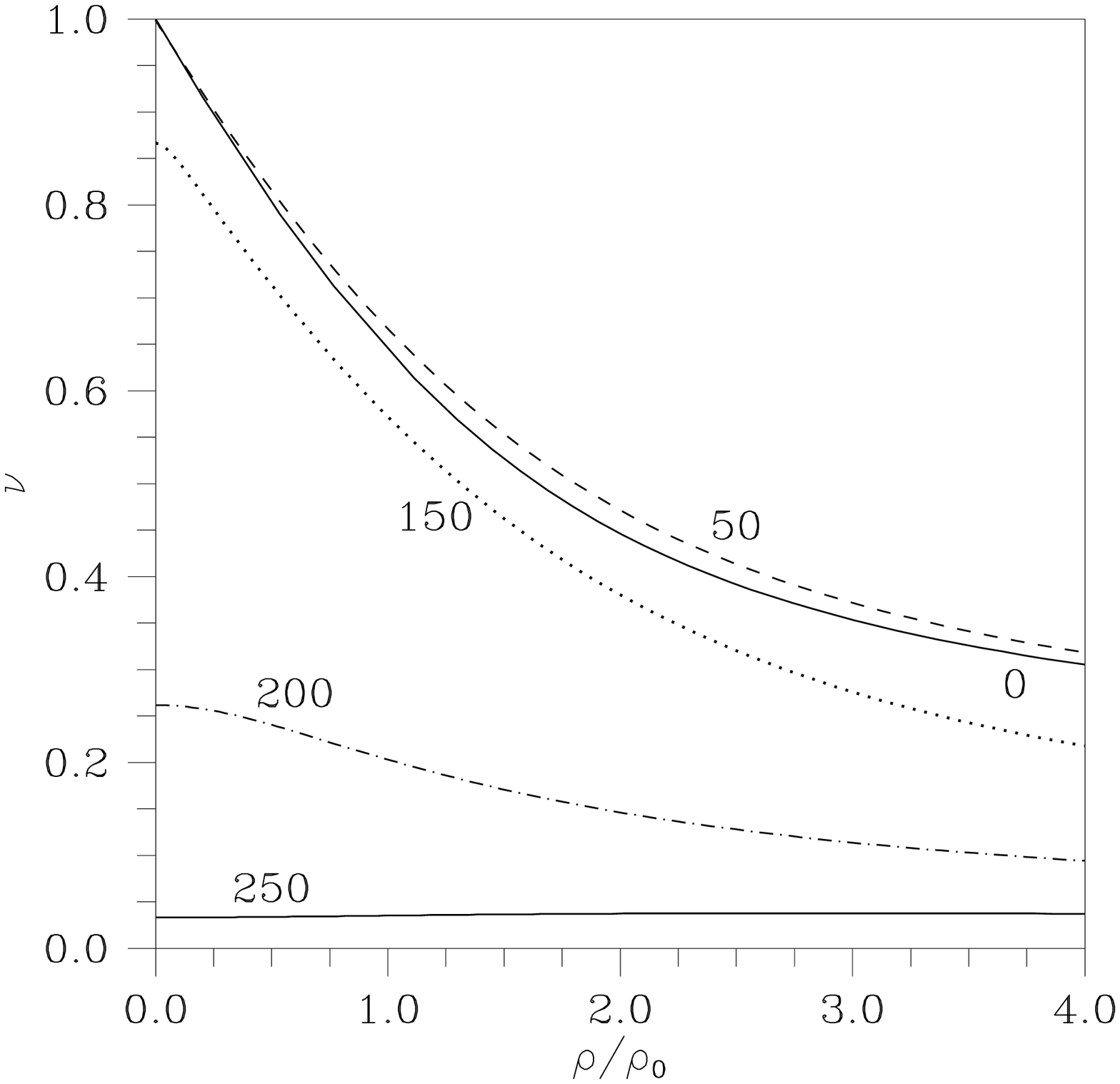}}
\noindent {\bf Figure 4.} The mean sigma field, 
$\nu=\bar{\sigma}/\sigma_0$, as a function of density for 
various temperatures (in MeV) with $\epsilon_1'>0$.
\end{figure}

\begin{figure}[t]
\setlength\epsfxsize{6.0in}
\centerline{\epsfbox{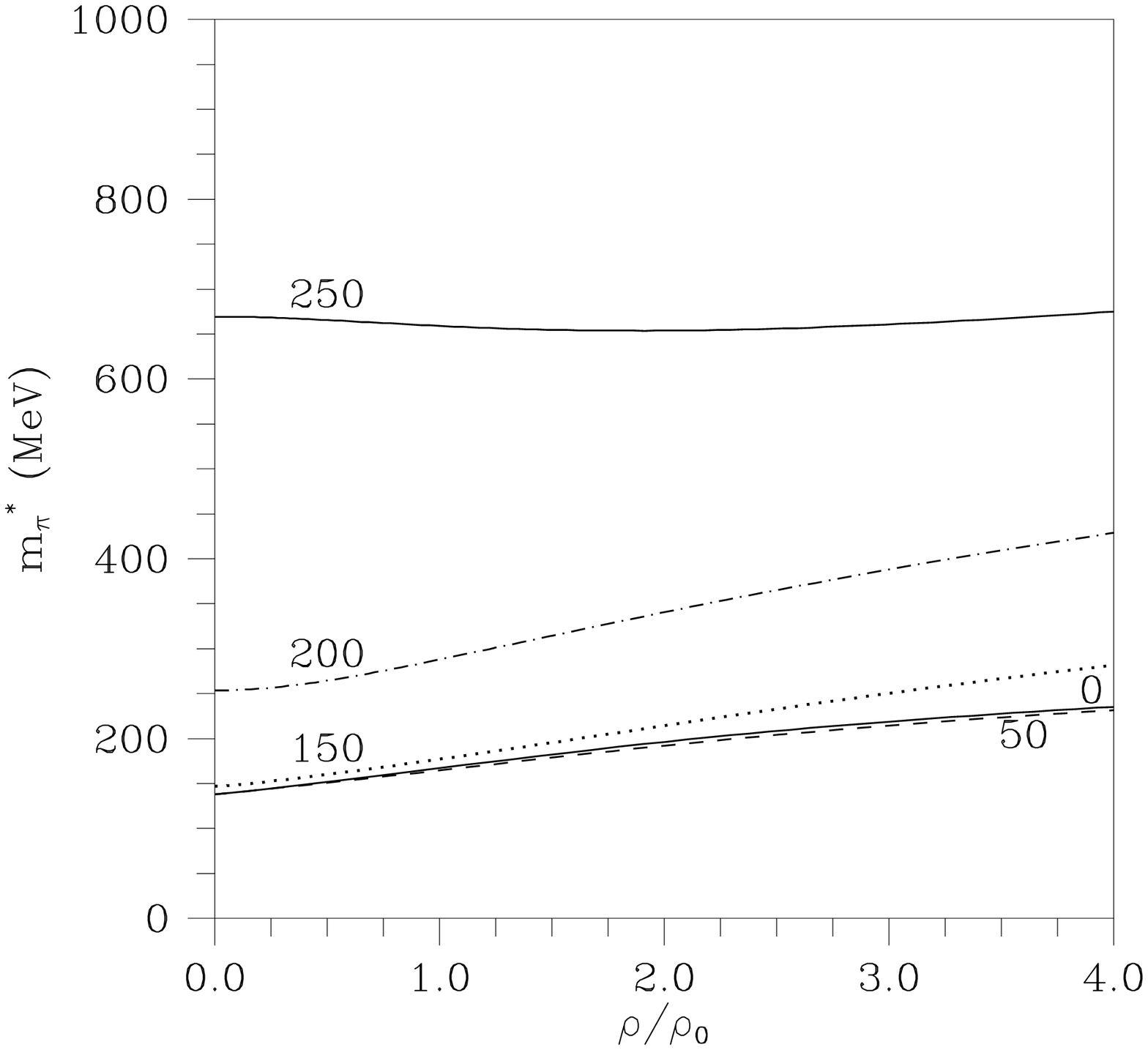}}
\noindent {\bf Figure 5.} The pion effective mass, $m_{\pi}^*$, 
as a function of density for various temperatures (in MeV)
with $\epsilon_1'>0$.
\end{figure}

\begin{figure}[t]
\setlength\epsfxsize{6.0in}
\centerline{\epsfbox{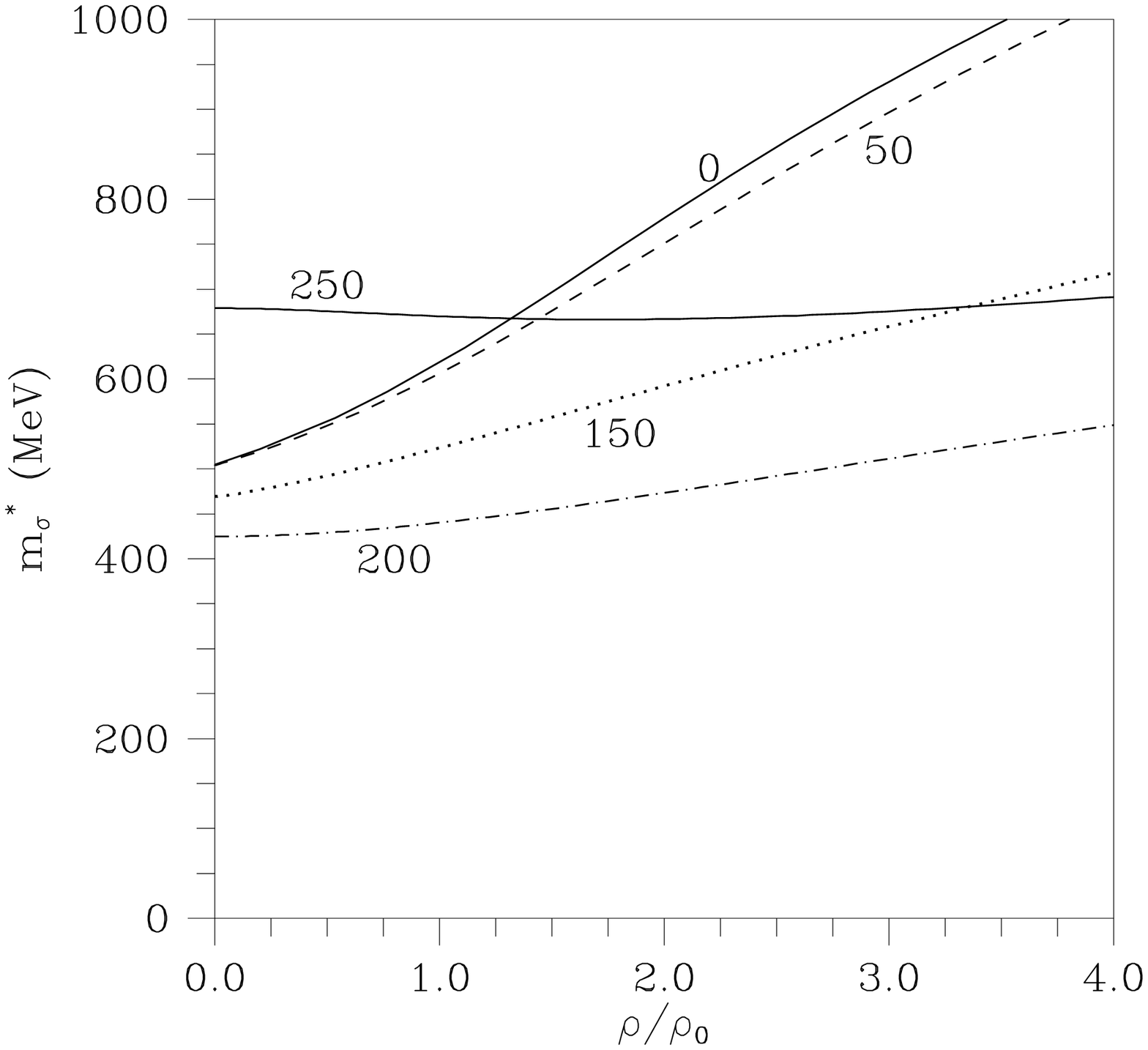}}
\noindent {\bf Figure 6.} The sigma effective mass, 
$m_{\sigma}^*$, as a function of density for various temperatures 
(in MeV) with $\epsilon_1'>0$.
\end{figure}

\begin{figure}[t]
\setlength\epsfxsize{6.0in}
\centerline{\epsfbox{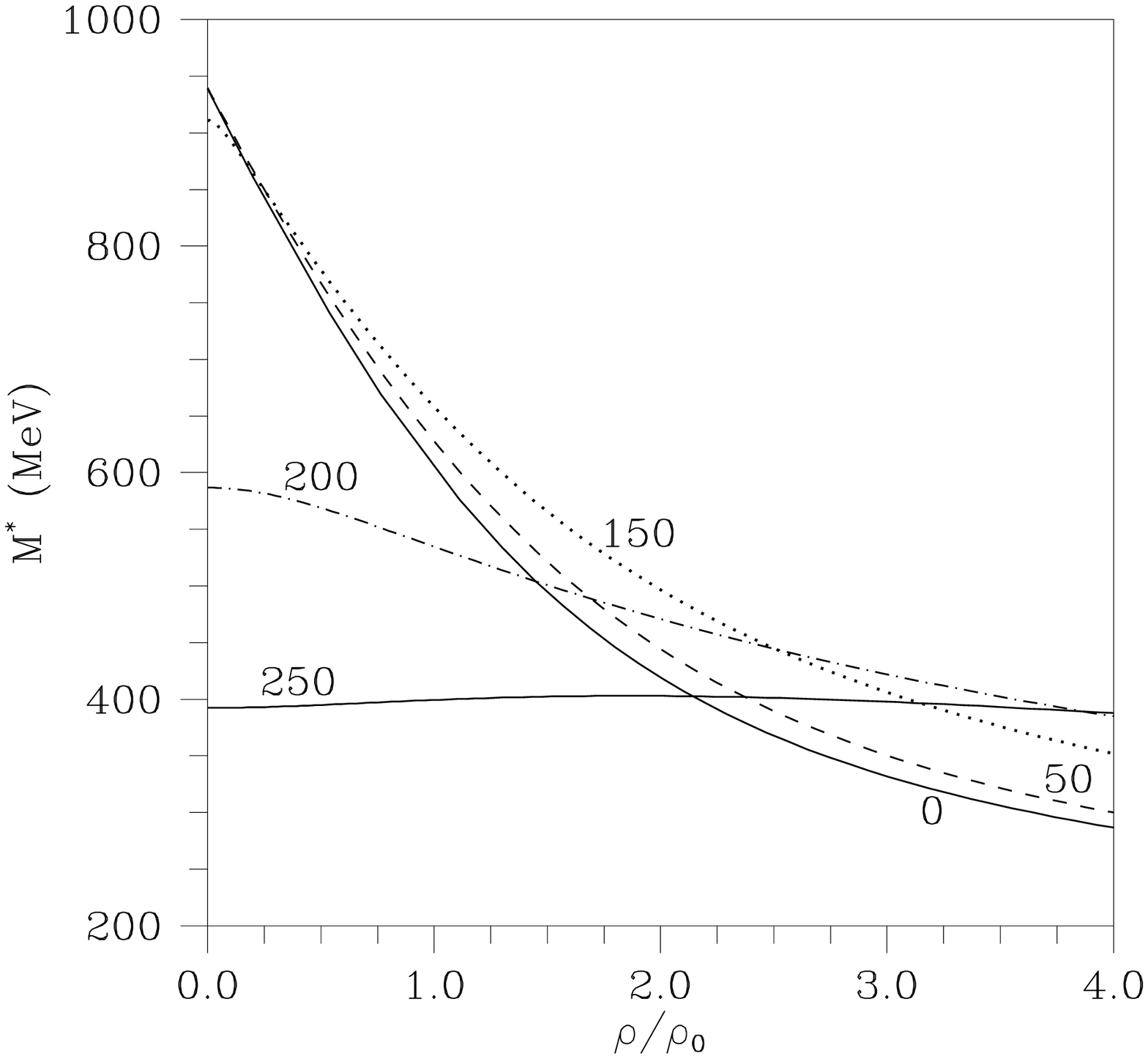}}
\noindent {\bf Figure 7.} The nucleon effective mass, $M^*$, 
as a function of density for various temperatures (in MeV)
with $\epsilon_1'>0$. 
\end{figure}

\begin{figure}[t]
\setlength\epsfxsize{6.0in}
\centerline{\epsfbox{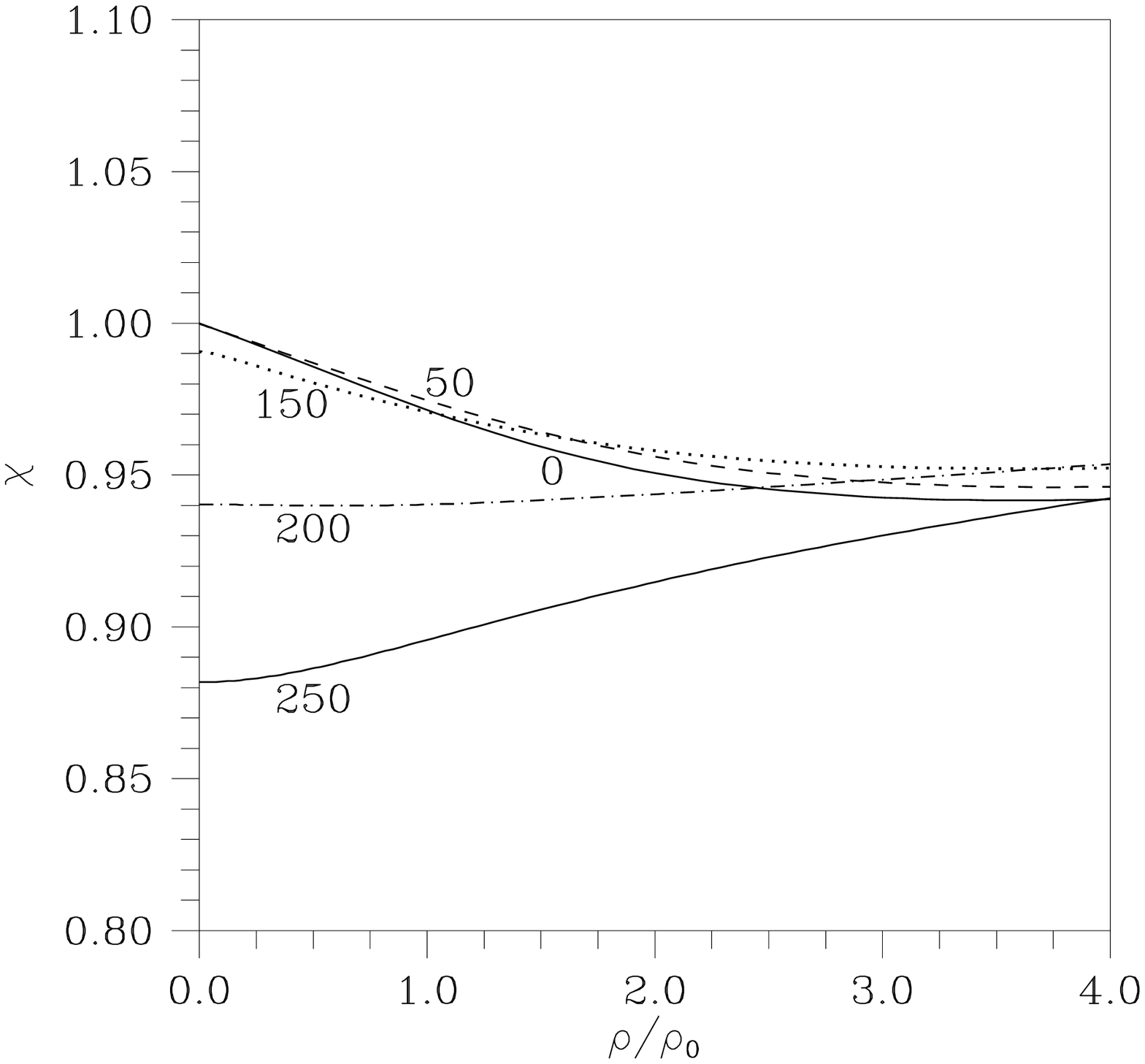}}
\noindent {\bf Figure 8.} The glueball mean field, 
$\chi=\phi/\phi_0$,
as a function of density for various temperatures (in MeV)
with $\epsilon_1'>0$.
\end{figure}

\begin{figure}[t]
\setlength\epsfxsize{6.0in}
\centerline{\epsfbox{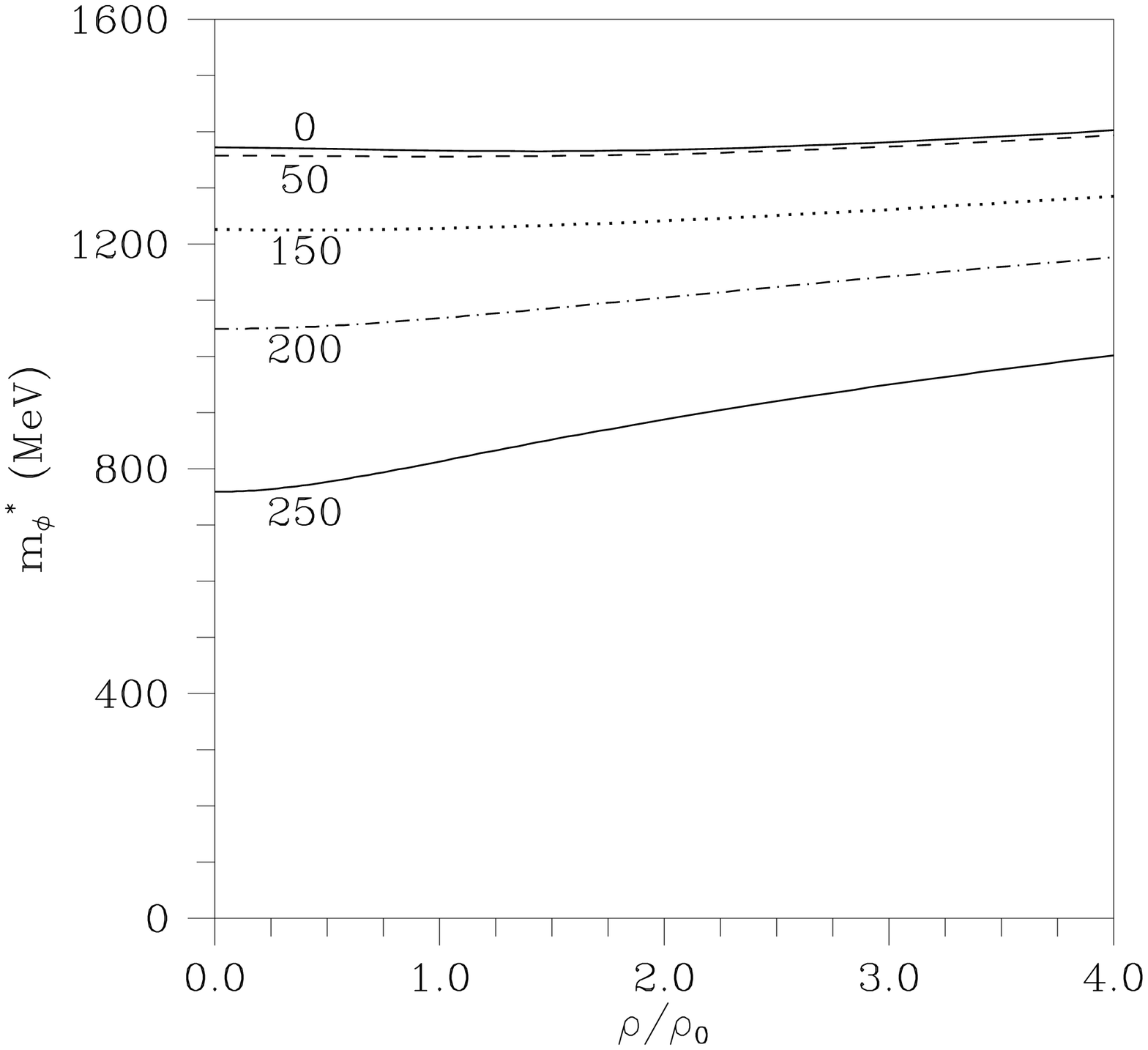}}
\noindent {\bf Figure 9.} The glueball effective mass, 
$m_{\phi}^*$,
as a function of density for various temperatures (in MeV)
with $\epsilon_1'>0$.
\end{figure}

\begin{figure}[t]
\setlength\epsfxsize{6.0in}
\centerline{\epsfbox{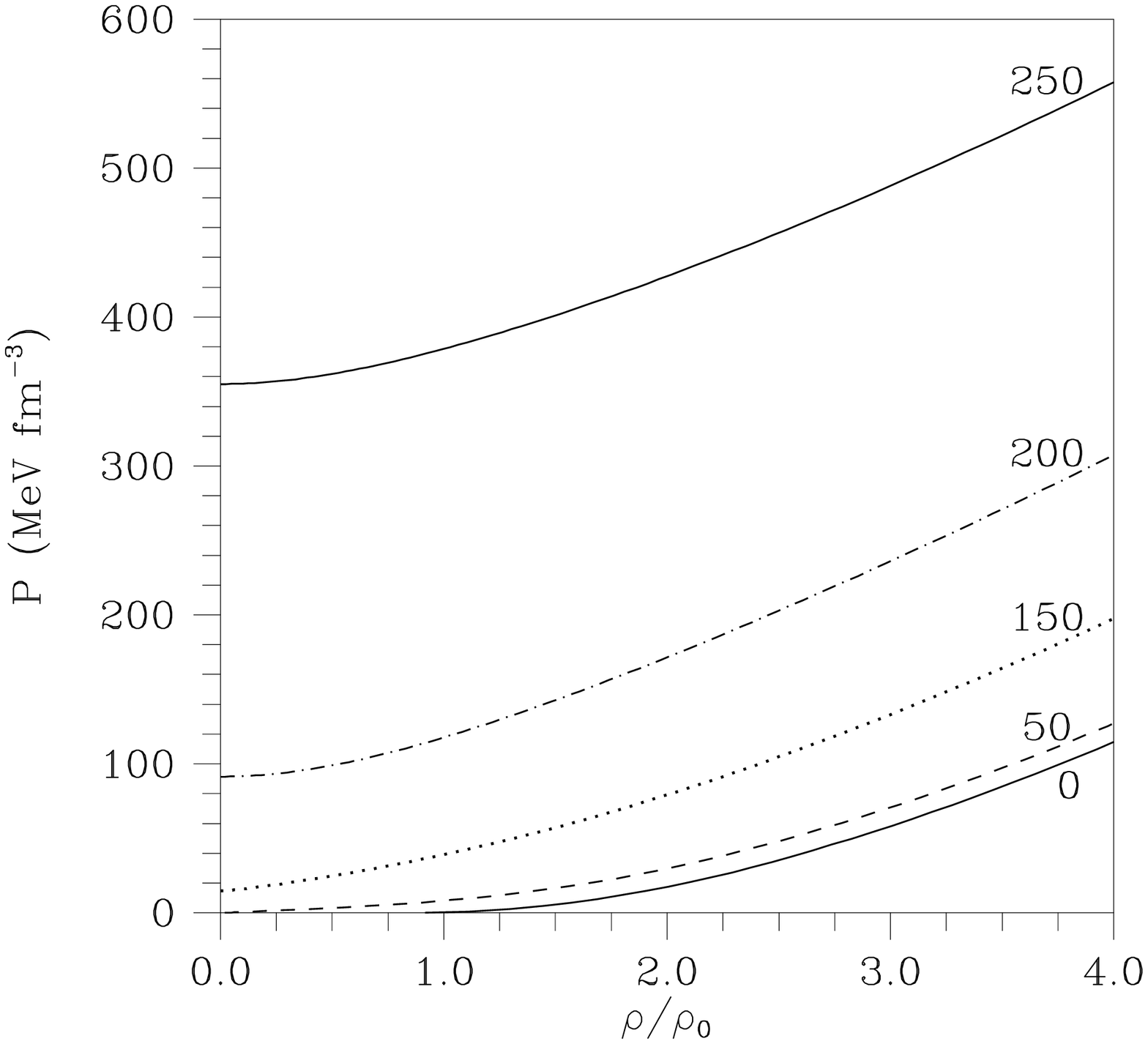}}
\noindent {\bf Figure 10.} The pressure as a function of 
density for various temperatures (in MeV)
with $\epsilon_1'>0$.
\end{figure}

\begin{figure}[t]
\setlength\epsfxsize{6.0in}
\centerline{\epsfbox{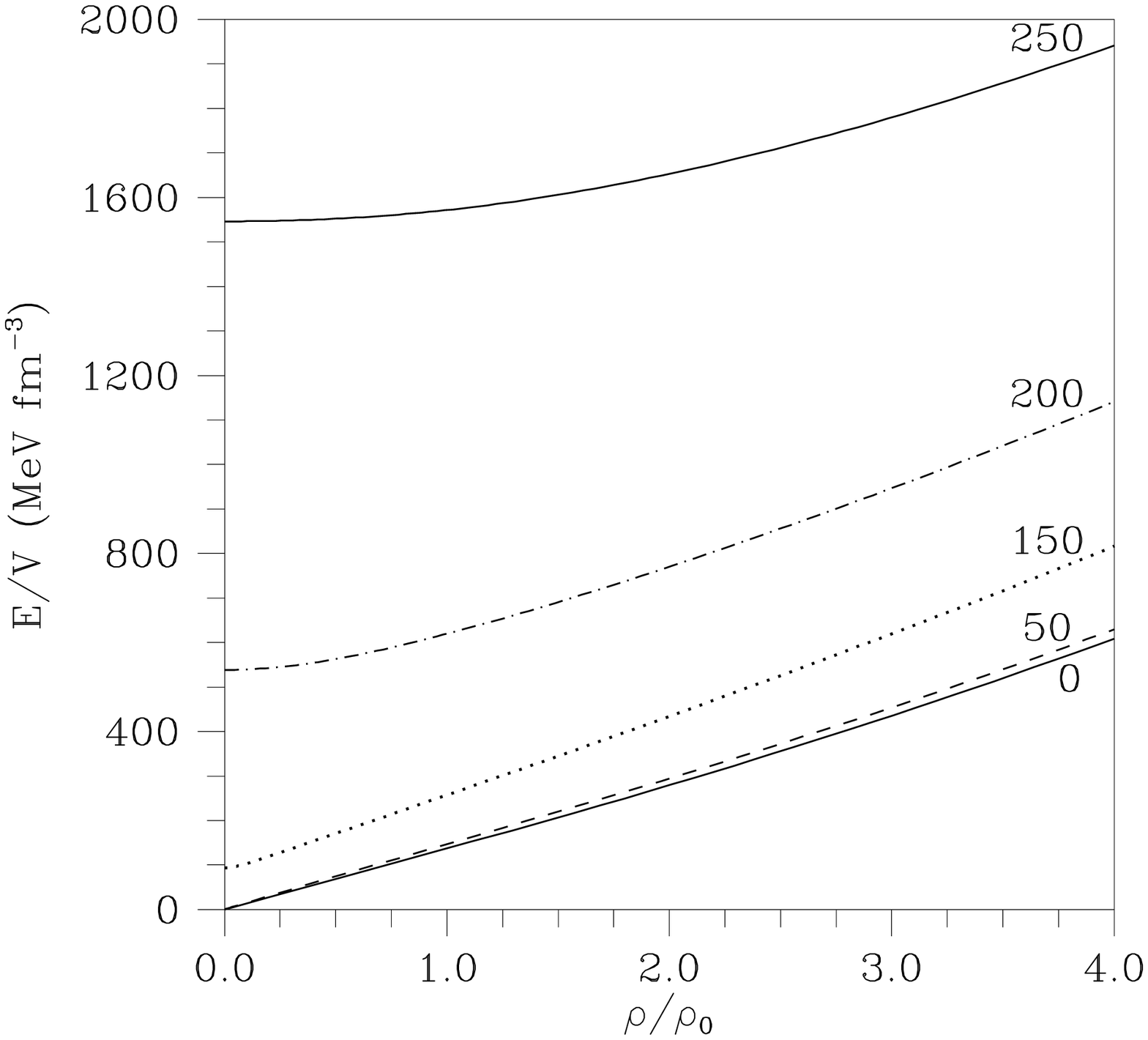}}
\noindent {\bf Figure 11.} The energy density as a function of 
baryon density for various temperatures (in MeV)
with $\epsilon_1'>0$.
\end{figure}

\end{document}